# Resource pricing in a Dynamic Multi-Commodity Market for Computational Resources

K. Abdelkader, J. Broeckhove and K.Vanmechelen

Department of Mathematics and Computer Sciences, University of Antwerp,
Antwerp, Belgium
Khalid.Abdelkader@ua.ac.be
Kurt.Vanmechelen@ua.ac.be
Jan.Broeckhove@ua.ac.be

*ABSTRACT*

*The adoption of market-based principles in resource management systems for computational infrastructures such as grids and clusters allows for matching demand and supply for resources in a utility maximizing manner. As such, they offer a promise of producing more efficient resource allocations, compared to traditional system-centric approaches that do not allow consumers and providers to express their valuations for computational resources. In this paper, we investigate the pricing of resources in grids through the use of a computational commodity market of CPU resources, where resource prices are determined through the computation of a supply-and-demand equilibrium. In particular, we introduce several categories of CPUs characterized by their execution speed. These differ in cost and performance but may be used interchangeably in executing jobs and thus represent so-called substitutable resources. We investigate the performance of the algorithms for computing the supply-and-demand equilibrium in this multi-commodity setting under dynamically varying consumer and provider populations.*

## 1. INTRODUCTION

The key function of a resource management system (RMS) is to efficiently allocate tasks to available resources. An efficiently functioning RMS is particularly important in the case of large-scale heterogeneous infrastructures that span many administrative domains such as grids or public resource computing networks. In many systems that are in use today, resource control and scheduling are often driven by system-centric objectives such as optimizing system throughput. However, the large user base of such infrastructures includes strongly varying requirements and preferences with respect to resource allocations. It is therefore important that there be a shift towards a user-centric approach that allows the individual expressions of the values of tasks and resources by consumers and providers, and has the resource management system take those valuations into account in allocation and scheduling decisions [1].

The introduction of market-based principles into resource management addresses that issue. It involves the definition of a computational market in which grid users and providers interact. Prices are used to signal supply and demand of resources. They reflect the users' valuations of the resource usage at particular points in time and signal the abundance or scarcity of various types of resources. Providers sell access to computational resources to consumers in a market that can be organized in different ways, e.g. posted-price markets, different types of auctions or commodity markets [2].

In this contribution, we consider the commodity market model with centralized equilibrium pricing [3]. It corresponds to a vision of the grid where applications can treat computational resources as interchangeable and not as specific machines and systems. Market participants express their valuation for resources as a function of their price. That means that for each price level the consumers, i.e. users wanting to obtain resources to have their jobs executed, indicate how many resources they will acquire. Similarly, for each price level the providers, organizations or individuals bringing their resources into the grid, indicate how many of their resources they will make available. The market sets the actual price by computing the price at which supply equals demand.

The pricing scheme based on supply and demand equilibrium in a computational commodity market has been investigated by [4]. It has been found to function well for static grids, i.e. grids with a static population of consumers and providers. Under these circumstances, the determination of the equilibrium price under variations of e.g. job load or disposable budget is fast and stable. This observation has been extended by [5] to dynamic infrastructures and by [6, 7] to large scale systems.

In this contribution, we extend this work to several substitutable resources. We introduce several categories of CPUs characterized by execution speed. These differ in cost and performance but may be used interchangeably in executing jobs and thus represent so-called substitutable resources. We investigate, using the Grid Economics Simulator [8], whether the algorithms for computing the supply-and-demand equilibrium, function in this context.

This paper is organized as follows. Section 2 discusses the commodity market model which includes the consumer, provider and pricing models used in the market. Section 3 provides a brief description of the Grid Economics Simulator (GES) that we have used in this study for obtaining empirical results. Section 4 describes the results obtained under different simulation scenarios, after which we formulate our conclusions.

## 2. COMMODITY MARKET MODEL

Modeling a resource market involves modeling the resources that are traded, the operation of the market (in particular the pricing mechanism), and the behavior of the market participants, i.e. the providers and consumers. With the exception of the resource model, which we extend in this work to include several substitutable categories to model a range of CPU performances, all these models have been explained in detail in [4, 6, 7]. We limit ourselves to a brief description for those models here.

### 2.1. Resource Model

A single type of resources, namely $CPU_s$ has been used in our investigations to date. In order to represent the diversification in CPU performance, we introduce a number of CPU categories, indexed by $i$, which we refer to as $CPU_i$. Their performance is expressed normalized with respect to that of $CPU_1$ that acts as a reference. In the present paper we have, somewhat arbitrarily, chosen performance ratios $r_i$ to be linear: CPUs from category $CPU_2$ execute a task twice as fast as those from category $CPU_1$, CPUs from category $CPU_3$ three times as fast and so on. In a real-world deployment, the determination of these ratios can be adapted to the variance one observes in the actual CPUs that are integrated in the grid infrastructure, through clustering methods. These categories constitute substitutable commodities, i.e. jobs can execute on both, but consumers will value them differently.

### 2.2. Job Model

As a consequence of limiting our resource model to CPUs, we also model a job as a CPU-bound computational task. Every job is characterized by a normalized running time. This time corresponds to the time it takes for the job to run on a CPU with a performance ratio of 1. However, in our algorithms we do not assume that the consumer has knowledge of this running time.

Jobs are taken to be atomic in the sense that they are always allocated to a single CPU and are non-preemptable. The dispatch of a job to a CPU is affected immediately after the necessary resource has been acquired. Initially, every consumer's queue has a number of jobs in it; with a certain probability jobs are added to the queue at every simulated time step or at periodic intervals. The former type of job injection models a continuous background load, the latter the occurrence of peak loads.

## 2.3. Consumer Model

The consumers in the computational commodity market are modelled as agents that act on behalf of grid users. Each consumer has a queue of computational jobs that need to be executed and for which resources must be acquired from providers through participation in the market. The price a consumer is willing to pay depends on the valuation for its jobs. More specifically, a consumer is characterized by a vector $v$ of valuation factors. The $i^{th}$ component of this vector ($v_i$) is used in conjunction with the performance ratio $r_i$, to normalize the price level for CPU category $i$ (denoted by $p_i$) that is observed in the market. This allows the consumer to compare price levels of the different CPU categories based on its individual preferences, and make decisions on formulation of demand for the different categories accordingly. The normalized price for CPU category $i$, is then given by equation 1.

$$p_i^{norm} = \frac{p_i}{r_i * v_i} \qquad (1)$$

The r.h.s reflects the price normalized to unit performance and factors in the consumer's valuation for the category. The use of the $v_i$ term provides a simple abstraction for the complex logic a consumer might follow to prefer one CPU category over another. An example of such a logic whereby a consumer is willing to pay more than double the price for a CPU of category 2, which is only twice as fast as one of category 1, is the following. Suppose the consumer has a job graph that includes a critical path and that a scheduling strategy is used to optimize for total turnaround time. Such a consumer would be willing to pay more than the nominal worth of a CPU of category 2 for allocating jobs on the critical path, as they have a potentially large effect on turnaround time.

Each consumer is provided with an initial budget and an additional periodically replenished allowance. In every simulation step, consumers are charged with the usage rate prices for all grid resources that are currently allocated to their jobs. In the market, contracts are established between providers and consumers for the full duration of a job. That is, $p_i$ denotes the price that will be charged to the consumer for the entire duration of the job, if it is run on a resource from category $i$.

Consumers do not attempt to save up credits, but try to use their entire budget. However, expenditures are spread out evenly across the allowance period. This is done because we assume that consumers do not have reliable estimates of the running time of their jobs. Therefore, we need to prevent consumers from agreeing to a cost level that would not be sustainable for them over the entire allowance period. Under these modeling decisions, consumers then formulate demand in the category with the lowest value for $p_i^{norm}$. The volume of this demand is bounded by the consumer's budgetary constraint.

## 2.4. Provider Model

Every provider hosts a number of CPUs in each category that can be supplied to the market. Once a resource is allocated to a job, it remains allocated until the job completes. Also, the market price at the time the resource is sold will be charged as a fixed rate to the consumer for the duration of the job. This approach is consistent with the fact that we do not assume a prior knowledge of a job's running time.

For a given price vector $p$ that has a price component for each CPU category in the market ($p_i$), providers have to determine how many CPUs they are willing to sell for each CPU category. To do so, a quantity $MPR_i$ is calculated. This denotes the "mean provider revenue per time step and per resource" for category $i$. The $MPR_i$ reflects the price the provider was able to obtain, on average, in the past for that category.

Given the price $p_i$ for resources of category $i$ and a number of available resources $PC_i$ in that category, the provider indicates a willingness to supply a number of resources in category $i$, as given by equation 2.

$$Supply_i = PC_i * \min\left(1.0, \frac{p_i}{MPR_i}\right) \tag{2}$$

That is, at a price that exceeds average past revenue, all resources are made available; at a price below that level, a share proportional to the ratio is made available. Providers thus limit their supply to the market in order to keep prices high, thereby trying to maximize revenue.

However, the fewer resources are sold, the lower the $MPR_i$ becomes, and this will in turn increase the number of resources offered at price level $P_i$. The duration of the history period used to determine the $MPR_i$ has a significant impact on the speed with which the provider reacts to market circumstances. At one extreme, when the window reduces to the current time step, $MPR_i \rightarrow p_i$ and providers make all resources available. At the other extreme, when the window includes all previous time steps, the $MPR_i$ becomes rigid and short term evolution has little impact. The length of the window can be used to encode the provider's reluctance to react to short term change in price levels.

### 2.5. Resource Pricing

The market mechanism that we adopt is a spot market, with a dynamic price for resources, adjusted to bring about market equilibrium in order to match supply and demand. In this setting, prices play the role of a communicator of complex provider and consumer valuations of the resources [9]. The market brings together all parties to quote their supply or demand for a range of prices for each CPU category. This information is used (after pre-processing and smoothing related to the fact that we are dealing with discrete resource units), to define an excess demand surface i.e. the difference between current demand and supply as a function of the price vector.

Figure 1 represents such a surface for a commodity market with two CPU categories. The excess demand function is denoted by $\xi$. Thus, $\xi_i(p)$ is the excess demand for the ith good at price vector $p$. We use the familiar Euclidian vector norm as a correctness measure for a given price vector $p$ (cfr. equation 3).

$$|\xi(p)| = \sqrt{\sum_{i=0}^{n} \xi(p_i)^2} \tag{3}$$

A norm greater than zero points to an inefficiency in the sense that the suggested market price p, does not lead to a market equilibrium. In such a case, resources are not necessarily allocated to consumers that value them the most or might remain unallocated.

The market equilibrium point is the zero of the excess demand surface and fixes the price at which the market will trade at that point in time. The algorithm that we use for computing this equilibrium is based on Smale's method [10], with modifications and extensions [4]. At its core, this algorithm follows a Newton-like approach, using information from the partial derivatives on the excess demand function ξ, to iteratively adjust the price levels. We refer to this algorithm as ESGN (Extended Smale Global Newton). The ESGN algorithm is further combined with an augmented Lagrangian pattern search [11]. This algorithm is tailored towards optimization problems with nonlinear objective functions or constraints. It does not rely on accurate derivative information and is therefore more robust with respect to discontinuities in the excess demand function. We have used the implementation of this algorithm that is delivered by the Matlab Genetic Algorithm and Direct Search Toolbox.

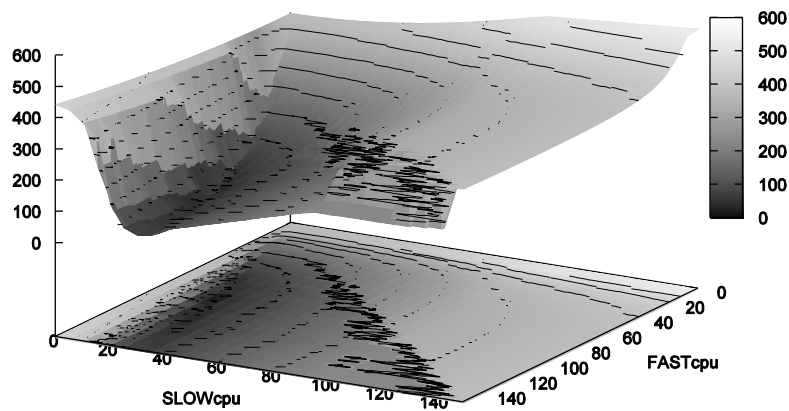

Figure 1. Sample excess demand surface for a commodity market with two CPU categories, the Z-axis denotes $|\xi(p)|$

The resulting optimization process first invokes the ESGN algorithm. When ESGN does not arrive at an equilibrium price, and the associated value of $|\xi(p)|$ exceeds a preconfigured threshold, the pattern search algorithm is invoked to further improve $|\xi(p)|$. We allow this optimization process to iterate and compute a new price vector $p'$, as long as constraints are fulfilled with respect to:

- The current value of $|\xi(p)|$
- The number of times the optimization process has already iterated
- The size of the improvement made by invoking the pattern search algorithm

The threshold values for these constraints can be configured on a case by case basis. In the remainder of this paper we will use the following values:
- $|\xi(p)| < 80$
- Number of iterations < 10
- $|\xi(p)| - |\xi(p')| > 1$

## 2.6. Grid Membership Dynamics of Consumers and Providers

We use the term "grid" in a broad sense as a large network of independent provider and consumer nodes cooperating to achieve the completion of computational tasks. We interpret this as an inherently dynamic setup, with nodes leaving and joining or rejoining the grid, for whatever reason.

We use a simple, straightforward model for the dynamics of peer participation or churn in the network. There is a pool of potential providers and a pool of active providers. A given rate governs the departure of providers from the former to join the latter. Similarly active providers depart at a given rate. In that event, we let the provider's CPUs finish processing the jobs currently executing on it. Only then does the provider enter the pool of potential providers. This state is referred to as the de-activated state for the provider. The size of the pools and the balance of the departure rates determine the evolution in time of the number of active providers.

An analogous mechanism is in place for consumers. Active consumers have jobs and participate in the market. There is a pool of potential consumers that can enter the market. State changes are determined by comparing randomly drawn numbers against pre-set thresholds. If a consumer is deactivated, it will arrive in the pool of potential consumers after its currently running jobs have finished. Upon deactivation, a consumer retains the jobs in its queue.

## 3. THE GES SIMULATOR

Our study uses the Grid Economics Simulator (GES) [8] to evaluate the performance and behavior of the proposed commodity market model. GES is a Java-based simulator that supports both discrete-time and discrete-event simulation, that is aimed at the study of various economic resource management approaches for large-scale distributed infrastructures. It focuses on analyzing the resource management system's ability to efficiently organize a resource market. GES also supports non-economic approaches in which case aspects such as billing and pricing are excluded from the simulation process.

Figure 2 gives a general overview of the GES architecture in terms of layers and components. Inside the **Core** layer, the **domain** layer contains base classes for domain entities such as **Consumer**, **Provider**, **GridResource**, and **GridEnvironment**. Support for traditional forms of resource management is provided through the **Non-Economic** layer and for market-based resource management in the **Economic** layer. On top of each of those one finds specific resource management systems. Outside the **Core** one finds layers dealing with configuration of simulations and with distributed processing of simulations based on technologies such as **Jini**, Sun Grid engine (**SGE**) or directly through system-to-system secure shell connections (**SSH**). This can of course be extended to other distribution platforms. A more detailed technical overview, and a comparison with similar simulators can be found in [8].

In this contribution, we have used the discrete-time capabilities of the simulator. In the discrete-time model, time advances in simulation steps. During each step a number of actions are taken. Each simulation step in the commodity market consists of the following actions:

1. Determination of the set of active consumers and providers for this step of the simulation.
2. The job queues of the active consumer's are updated.
3. If there are non-occupied resources in the market, equilibrium price levels are established for these resources.
4. Resource trading occurs at these price levels.
5. The available budget of all active consumers is updated. Payments for running jobs are made to the provider accounts.
6. The remaining runtime of all active jobs is updated and all jobs that have finished their execution are removed from the system.
7. The availability state of the provider's resources is updated.

GES has a user interface that consists of a set of utilities and controls to easily configure market scenarios and efficiently analyze and monitor the system's behavior. It also allows a very flexible composition of views on the various simulation data and results.

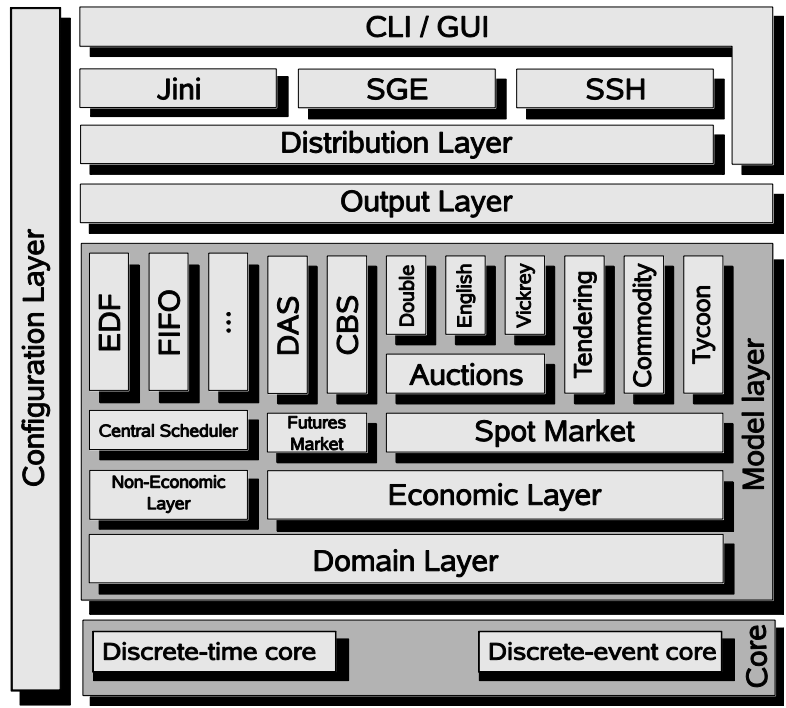

Figure 2. Overview of the architecture of GES

## 4. RESULTS

Using the GES simulator we verify that the commodity market operates correctly, i.e. that correct equilibrium prices can be determined at each simulated time step, while increasing the number resource categories in the system in successive simulation scenarios. Categories were configured with linearly increasing speed factor, i.e. the fourth category has a performance ratio $r_i$ of 4. For the purpose of analysis and comparison we report here on the results of a scenario with three and a scenario with six CPU categories. We have also explored similar scenarios with one, two, four and five categories. The results in those cases are similar to those reported in this paper and support its findings.

### 4.1. Simulation Scenarios

The key GES parameters that apply in the simulated scenarios are listed in Table 1. When a range is indicated, the parameter is determined by a uniform distribution in that range.

The number of consumers and providers in the simulation and the departure rates for pool transferrals are listed in Table 1. The departure rates are identical for both active and potential pools leading to a regime without long term shifts in the number of consumers and providers. All consumers and providers start out in the active pool. And as such, a higher load will be placed on the system in the beginning of the simulation. A small background load is generated at each time step by having each consumer, with a probability of 15%; add a new job to its queue. A periodic load is generated by injecting a large number (in the range {1, 2, ⋯, 150}) of jobs in the consumer's queue at every 50 time steps. The latter is done in order to probe the stability of price levels and gauge the market's response to the load spike. Jobs take {2, ⋯, 10} timesteps to process on the reference CPU with a performance ratio of 1. Budgets are replenished, also at every 50 time steps, with an amount equal to the initial budget. Consumers have uniformly distributed valuation factors for the different CPU categories that vary between 1.0 and 1.5.

Table 1. Simulation parameters.

| Common parameters for all scenarios | Value |
|---|---|
| Initial size, active consumer pool | 2000 |
| Departure rate, active consumer pool | 0.1 |
| Initial size, potential consumer pool | 2000 |
| Departure rate, potential consumer pool | 0.1 |
| Initial size, active provider pool | 1000 |
| Departure rate, active provider pool | 0.1 |
| Initial size, potential provider pool | 1000 |
| Departure rate, potential provider pool | 0.1 |
| Job length in time steps | {2, 3, ⋯ , 10} |
| Job injection period in time steps | 50 |
| Number of jobs per injected | {1, 2, ⋯ , 150} |
| Probability of new jobs per time step | 15% |
| Initial budget $B$ | [50.000,125.000] |
| Budget replenishment period | 50 |
| Budget amount replenished | $B$ |
| Valuation factor $v_i$ for each category | [1.0, 1.5] |
| **Scenario with three categories** | |
| Number of CPUs per provider in $CPU_3$ | {1, ⋯ , 10} |
| Number of CPUs per provider in $CPU_2$ | {1, ⋯ , 15} |
| Number of CPUs per provider in $CPU_1$ | {1, ⋯ , 30} |
| **Scenario with six categories** | |
| Number of CPUs per provider in $CPU_6$ | {1, ⋯ , 3} |
| Number of CPUs per provider in $CPU_5$ | {1, ⋯ , 3} |
| Number of CPUs per provider in $CPU_4$ | {1, ⋯ , 4} |
| Number of CPUs per provider in $CPU_3$ | {1, ⋯ , 5} |
| Number of CPUs per provider in $CPU_2$ | {1, ⋯ , 7} |
| Number of CPUs per provider in $CPU_1$ | {1, ⋯ , 12} |

As indicated above, we intend to compare scenarios with different numbers of CPU categories. However, in order to make these scenarios comparable, we need to take care and ensure that the total processing capacity per time step that is potentially available in the market remains the same across the different scenarios. This total processing capacity is given by multiplying the average number of CPUs hosted per category with the performance ratio for that category and summing over the categories.

Figures 3 and 4 show the results of the three category and six category scenarios respectively. The top panels indicate the price evolution for each CPU category, and the bottom panel the utilization for each of the CPU categories. The Figures show that in both scenarios, the market succeeds in dynamically pricing resources in response to the varying demand and supply levels. In each period of 50 simulation steps, prices peak as jobs are injected in the consumers' queues and congestion ensues. As congestion eases of (which can be deduced from the bottom panel in each figure), competition in the market for resources is reduced and subsequently, lower price levels are registered. Note that we have chosen this scenario because of the large fluctuations in demand which requires the market to constantly adjust price levels to the new state of the environment.

As can be seen from the graph which shows the utilization in each CPU category, utilization levels for categories with a high performance ratio vary more widely than those for lower performance categories. This is partly due to the fact that there are fewer resources available in the system for the high performance categories. This is a consequence of our choice to distribute the available processing capacity almost evenly over the different CPU categories. As a result, an unmatched resource in a high performance category has a higher impact on the average utilization for that category.

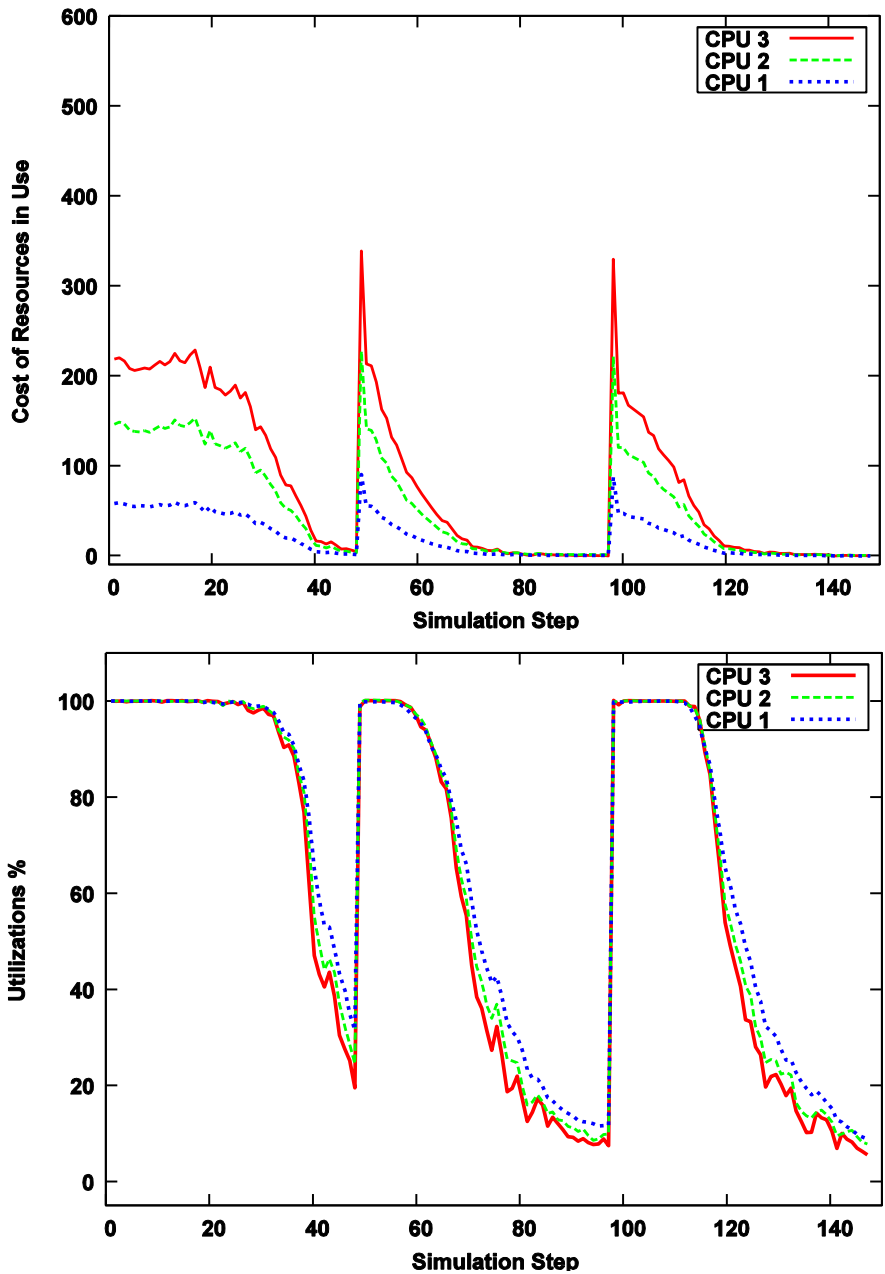

Figure 3. Results for the three-CPU-categories scenario. Price evolution (top panel) and utilization (bottom panel) for each of the three CPU categories.

In both scenarios, the market is able to quickly adapt to a demand shock in the environment, as evidenced by a reconfiguration of the price levels at steps 50 and 100. Note that the system's behavior during the initial job peak at simulation step 1 enfolds differently from those in steps 50 and 100. Prices and utilization levels remain at a high level for a longer period of time compared to the other two periods of peak load. This is caused by the fact that we introduce a higher load at the onset of the simulation by configuring each consumer to start out in the active consumer pool. As we can see from the figure, this is correctly reflected by the market's outputs.

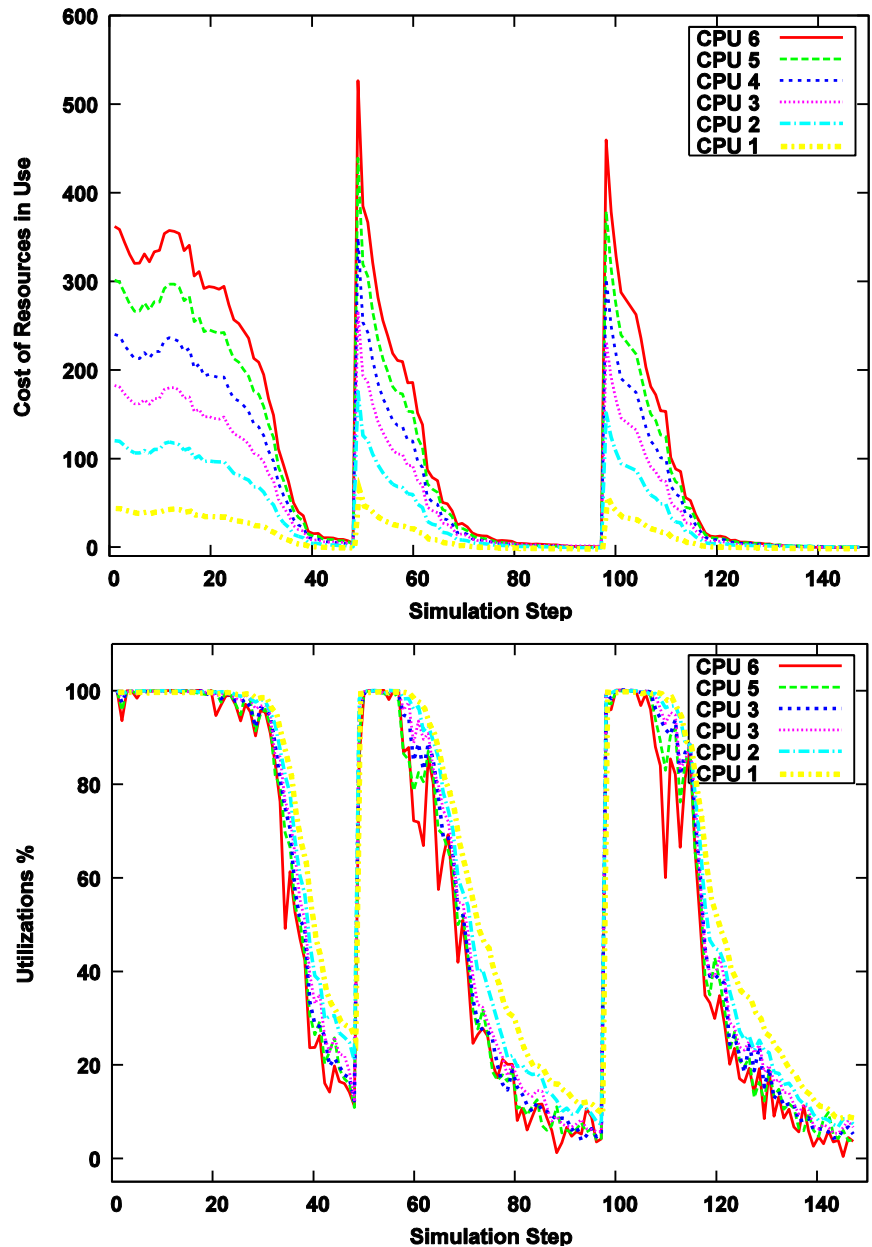

Figure 4. Results for the six-CPU-categories scenario. Price evolution (top panel) and utilization (bottom panel) for each of the six CPU categories.

Figure 5 shows that the consumer population as a whole spends a similar amount of money during the different steps in the simulation, irrespective of the number of CPU categories that are defined in the market. This supports the fact that on average, revenues for providers are not affected by introducing more CPU categories in the given simulation setup.

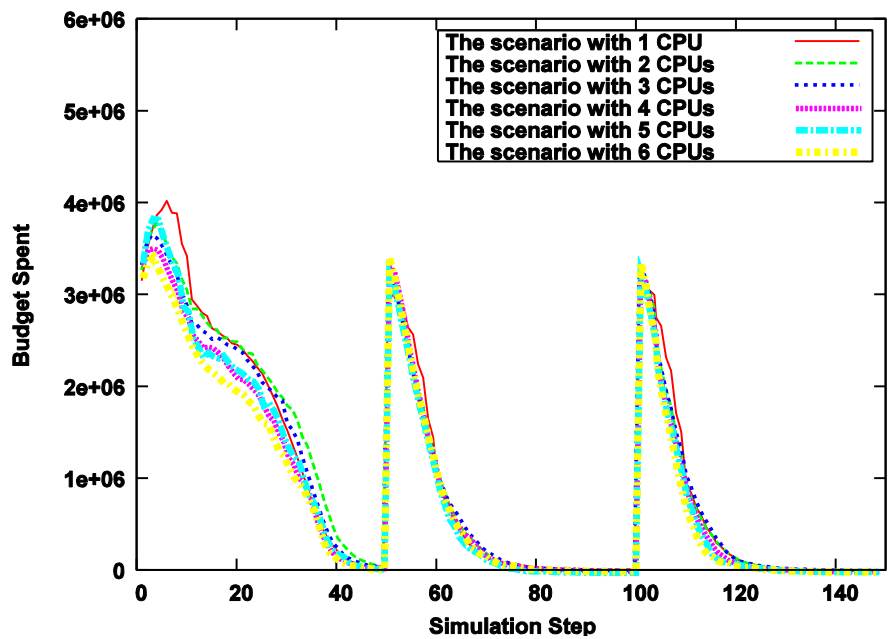

Figure 5. The total amount of budget spent on resources during the simulation for scenarios with differing numbers of CPU commodities.

Table 2. Breakdown of the results in the six category scenario.

| Categories | $CPU_1$ | $CPU_2$ | $CPU_3$ | $CPU_4$ | $CPU_5$ | $CPU_6$ |
|---|---|---|---|---|---|---|
| **Mean price** | 14.65 | 39.30 | 59.68 | 78.84 | 99.09 | 119.51 |
| **Avg. utilization (%)** | 63.35 | 60.48 | 57.91 | 56.38 | 55.08 | 52.86 |
| **Avg. value of $\xi_i$** | -11.36 | -16.18 | -23.07 | -23.22 | -27.08 | -36.68 |

Table 2 lists average price, utilization and excess demand levels for each of the CPU categories in the scenario with six CPU commodities. We note that prices are set by the market in line with the performance of the different categories and the valuations for these categories among the market participants. Utilization levels are lower for more performant categories. This is due to the larger effect an unmatched resource has on this metric in a high performance category, as mentioned previously. The table also shows the average value of the excess demand function over the entire simulation, for each category. This data shows, that on average, resources are slightly overpriced leading to negative excess demand (i.e. excess supply). The oversupply is larger for high-performance CPU categories. Note that this number of unmatched resources is relatively low compared to the trade volume in each simulation step. During congested periods, more than a thousand resources are traded in each CPU category every simulated time step. This number gradually diminishes in line with the utilization levels shown in Figure 4. Nevertheless, it is interesting to note that pricing errors result in excess supply on average. One can argue that, in the context of resource allocation in computational infrastructures, excess demand is to be preferred over excess supply (which results in underutilization of the infrastructure, whereas excess demand results in an inefficient allocation of resources from the viewpoint of value maximization). Further research is required to analyse this bias and to possibly adjust the price optimization algorithm to prefer prices that lead to overdemand instead of oversupply in case an equilibrium price vector cannot be found.

Table 3. The average runtime of the price determination process in each simulation step, number of excess demand queries and runtime per simulation steps.

| # CPUs | Runtime (sec) | # Queries | 95% CI of $|\xi(p)|$ | Avg. $|\xi(p)|$ | Max $|\xi(p)|$ |
|---|---|---|---|---|---|
| 1 | 0.51 | 46 | [1.65,8.65] | 5.15 | 194.36 |
| 2 | 5.95 | 683 | [18.6,28.92] | 23.76 | 187.33 |
| 3 | 13.83 | 1467 | [31.67,46.85] | 39.26 | 363.71 |
| 4 | 24.91 | 2010 | [23.46,137.30] | 80.47 | 4316.86 |
| 5 | 39.06 | 2759 | [42.59,238.75] | 140.67 | 5844.37 |
| 6 | 71.64 | 4271 | [79.2,116.64] | 97.92 | 666.11 |

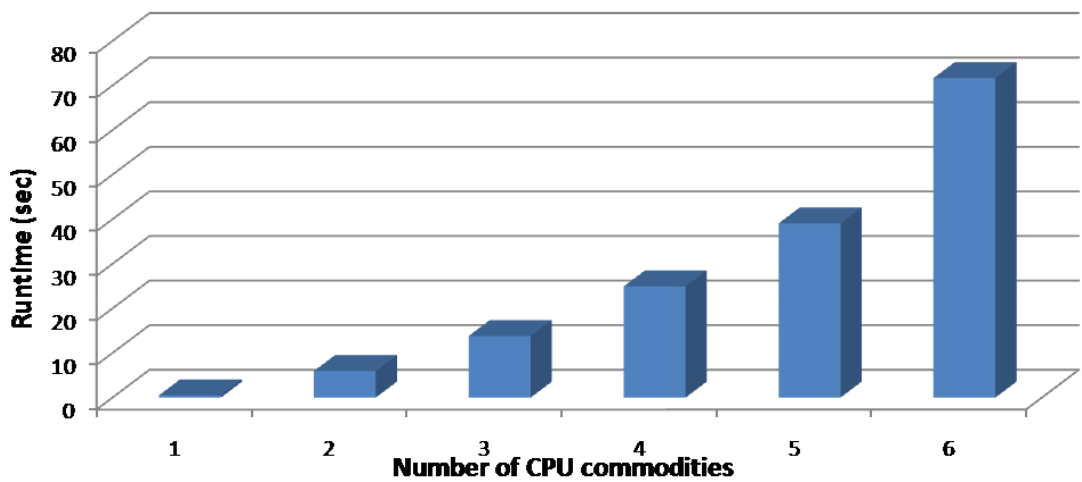

Figure 6. Average runtime of the equilibrium price search in each simulation step for different number of CPU commodities.

### 4.2. Computational Requirements

The introduction of more CPU categories in the market has the advantage of enabling market participants to more accurately formulate their requirements to the system and therefore express their valuations more accurately. As a consequence, more efficient resource allocations, from the viewpoint of value maximization, can be attained. However, introducing more categories also comes at the cost of increasing the complexity of the price determination process. The multi-dimensionality of the optimization problem of finding the equilibrium price significantly increases the runtime of this process. It also increases the value of the excess demand norm that is observed throughout the simulation, which leads to less accurate prices being set in the market. This is illustrated in Table 3. The diminishing price accuracy under a larger number of commodities is clearly reflected in the average, maximum and 95% confidence interval for the norm. We hereby note that the scenarios with 4 and 5 CPU commodities included a limited number of outliers that influence the results for both scenarios. If these are excluded from the data, the average norm and maximum norm for the scenario with four CPU categories become 51.5 and 599.6 respectively. The average norm and maximum norm for the five CPU category scenario then change to 70.75 and 546.5 respectively.

We observe a non-linear growth in both the runtime of the price optimization process and the number of excess demand queries that the optimizer needs during this process. Figure 6 illustrates this non-linear growth. All results were obtained on a single core of a Intel Quad Core processor with the cores running at 2.83 GHz. Although these runtimes do not preclude the use of a large number of CPU categories, care must be taken that the excess demand queries do not require (too much) network message roundtrips, which would lead to a too large overhead during the pricing process. This can be achieved by transmitting the consumer's bidding agent to the execution environment in which the price optimizer resides. In a Java-based setting this can easily be implemented through dynamic classloading. In the event that such a centralization would cause a bottleneck in terms of runtime performance, or memory use, one can opt to host the bidding agents on a number of nodes in a compute cluster.

## 5. FUTURE WORK

In future work, we plan to extend the consumer models to include time-varying valuations for jobs, dependent on their completion time. This also opens up the possibility to introduce more intelligent bidding behavior in the consumer that strategically times the acquisition of resources in the market, based on price levels and price level evolution, combined with the time-varying valuations for jobs. We are also interested in the use of trace data to simulate real-world platforms and workloads. Unfortunately, current workload archives do not include sufficient information in order to configure the budgetary capabilities of users, and their spending behavior. Finally, we will investigate the use of distributed and multi-core infrastructures to host the population of consumer and provider bidding agents and parallelize the calculation of total demand and supply in the system.

## 6. CONCLUSION

This paper investigates the functioning of a computational multi-commodity market that includes a number of substitutable resources representing CPUs with varying performance ratios. The results, obtained for different simulations, show that the core algorithm of the model, i.e. equilibrium price determination, is robust against the disruptive effects of the dynamic grid fabric even with several substitutable resources. Although the computational and communicative requirements of the price determination process do allow for the use of a significant amount of commodities in the market, they do increase non-linearly in this regard.

## ACKNOWLEDGEMENTS

Khalid Abdelkader gratefully acknowledges the support by the Libyan Government and the hospitality of the Department of Mathematics and Computer Science of the University of Antwerp.

**Authors**

Khalid Abdelkader is a PhD researcher and a member of the Computational Modeling and Programming (CoMP) group of the Department of Mathematics and Computer Sciences at the University of Antwerp (UA), Belgium. He completed his Master degree from the University of AGH, the Department of Computer Science, Krakow, Poland in 1999. His research interests include distributed systems, Grid Computing, and Grid Economics.

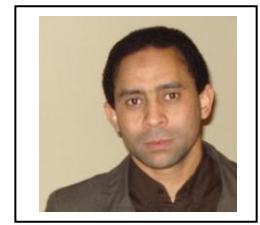

Prof. Dr. Jan Broeckhove is a professor in the Department of Mathematics and Computer Science at the University of Antwerp (UA), Belgium, and head of the Computational Modeling and Programming research group. He received his PhD, in Physics, in 1982 at the Free University of Brussels (VUB), Belgium. His current research interests include computational science and distributed computing, in particular cluster and grid computing.

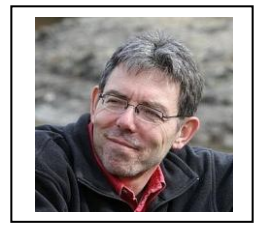

Dr. Kurt Vanmechelen is a post-doctoral researcher in the Computational Modeling and Programming (CoMP) group of the Department of Mathematics and Computer Science at the University of Antwerp (UA), Belgium. He received his PhD in Computer Science from the University of Antwerp in 2009. His research interests include Grid Economics in particular and Grid resource management in general.

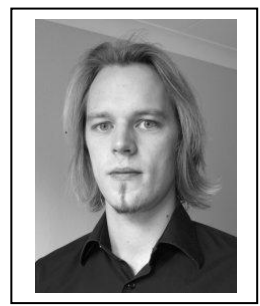